\documentclass[reprint,aps,pra,superscriptaddress]{revtex4-1}

\usepackage{graphicx}
\usepackage{xifthen}
\usepackage{xspace}
\usepackage{amsmath}
\usepackage{amssymb}
\usepackage{comment}
\usepackage[colorlinks, allcolors=blue]{hyperref}
\usepackage[all]{hypcap}
\usepackage[mathlines]{lineno}
\usepackage{braket}
\usepackage{wrapfig}
\usepackage{lipsum}
\usepackage{ulem}
\usepackage{units}

\newcommand{\pref}[2][]{\hyperref[#2]{\ref{#2}\ifthenelse{\isempty{#1}}{}{\sffamily(#1)}}}

\newcommand{\create}[2][n]{c_{#1\ifthenelse{\isempty{#2}}{}{,\text{#2}}}^\dagger}
\newcommand{\annihilate}[2][n]{c_{#1\ifthenelse{\isempty{#2}}{}{,\text{#2}}}}
\newcommand{\pnl}[1]{\textbf{\sffamily(#1)}}

\newcommand{\affuiuc}{\affiliation{Department of Physics, The University of Illinois at Urbana-Champaign, Urbana, IL 61801-3080, USA}}
\newcommand{\affcuhk}{\affiliation{Department of Physics, The Chinese University of Hong Kong, Shatin, Hong Kong, China}}

\begin{document}

\title{Counterdiabatic control of transport in a synthetic tight-binding lattice}

\author{Eric J.\ Meier}
\affuiuc

\author{Kinfung Ngan}
\affcuhk

\author{Dries Sels}
\email{dsels@g.harvard.edu}
\affiliation{Department of Physics, Harvard University, 17 Oxford St., Cambridge, MA 02138, USA}
\affiliation{Department of Physics, Theory of quantum and complex systems, Universiteit Antwerpen, B-2610 Antwerpen, Belgium}

\author{Bryce Gadway}
\email{bgadway@illinois.edu}
\affuiuc

\date{\today}

\begin{abstract}
\noindent Quantum state transformations that are robust to experimental imperfections are important for applications in quantum information science and quantum sensing. Counterdiabatic (CD) approaches, which use knowledge of the underlying system Hamiltonian to actively correct for diabatic effects, are powerful tools for achieving simultaneously fast and stable state transformations. Protocols for CD driving have thus far been limited in their experimental implementation to discrete systems with just two or three levels, as well as bulk systems with scaling symmetries. Here, we extend the tool of CD control to a discrete synthetic lattice system composed of as many as nine sites. Although this system has a vanishing gap and thus no adiabatic support in the thermodynamic limit, we show that CD approaches can still give a substantial, several order-of-magnitude, improvement in fidelity over naive, fast adiabatic protocols. 
\end{abstract}

\maketitle

\noindent In the adiabatic limit, high-fidelity transport and state-preparation can be achieved by slowly deforming a system's Hamiltonian, such that population remains in a given instantaneous eigenstate of the system, whose properties may be tuned through the Hamiltonian. If the Hamiltonian is changed too quickly, however, the state will be unable to track the instantaneous eigenstate, resulting in poor transport or unfaithful state preparation. The cause for this breakdown in adiabaticity is that when the Hamiltonian begins to change in time, it effectively acquires additional terms that can couple the different instantaneous eigenstates. These additional \textit{diabatic terms} are analogous to the \textit{inertial forces} that appear for systems represented in accelerating reference frames, such as when considering the motion of an egg relative to a supporting spoon as a child sets off on a spirited race.

In both the classical and quantum contexts, a range of optimization protocols can be developed that allow one to surpass the adiabatic limit~\cite{RevMod-Shortcut,EA-Widera,MachineLearningFoot,jensen2020crowdsourcing,driesshade2018,GRAPE,CRAB,Omran-GHZ,QAOA,Pagano-QAOA}.
When there exists a known Hamiltonian for the system being deformed, the diabatic terms that would induce transitions between instantaneous eigenstates can in principle be precisely determined and, with sufficient resources, directly counteracted by appropriate measures. Approaches along such lines, known as counterdiabatic (CD) techniques or shortcuts to adiabaticity, have been investigated in a physical context since the mid 2000s~\cite{RevMod-Shortcut, stareview}, beginning with works by Demirplak,~\textit{et\,al.}~\cite{demirplak2003, demirplak2005} and Berry~\cite{berry2009}. Theoretical formulations have been developed for small (two- and three-level) quantized systems~\cite{chen2010,Gerbier-CD-spinors}, as well as for transformations in continuous systems that have scaling transformations~\cite{delcampo2013,deffner2014,Deng-fermigas}. For the paradigmatic two-state problem, there have been demonstrations of counterdiabatic protocols across a wide range of experimental platforms, including atomic momentum states~\cite{bason2012}, nitrogen vacancy (NV) centers~\cite{zhang2013}, and superconducting qubits~\cite{zhang2017,Zhang2018,Wang2018}. More recently, CD protocols have even been demonstrated for the optimization of state transfer in 3-level NV centers~\cite{Awschalom-3states} and superconducting qutrits~\cite{SCqutrit}. The extension to larger discrete quantum systems has been significantly challenged due to the general difficulty in both calculating the proper counterdiabatic terms to be used~\cite{driespnas} and finding an experimental platform with the necessary level of control. 

Here, we extend the application of CD techniques to discrete many-level quantum systems. Using synthetic tight-binding lattices of laser-coupled atomic momentum states, we implement counterdiabatic driving protocols~\cite{driespnas} in a many-site system that has no adiabatic support in the thermodynamic limit. We demonstrate that CD driving protocols can lead to substantial improvements in state transformations across the lattice, such as in a multi-level adiabatic rapid passage as well as in the ability to prepare atoms in specific eigenstates delocalized across the synthetic lattices. The application of CD techniques in this context could lead to direct improvements in state transformations relevant to atom interferometry. Moreover, the native interactions in momentum-space lattices may enable future explorations into the influence of interactions on analog monopoles~\cite{Ho-monopole,Zhou-YangMonopole}, based on the connections between the physics of CD driving and topological invariants of the underlying parameter space~\cite{Gritsev6457}.

\section{Introduction}

\noindent Counterdiabatic approaches to surpassing the adiabatic limit are based on actively correcting for diabatic terms that emerge in the instantaneous reference frame when a Hamiltonian is rapidly deformed. The emergence of such terms is illustrated in Fig.~\pref[a]{CDIntro} for a canonical two-level system, having a bare Hamiltonian $H_0 = h_x \sigma_x+(V\!(\tau)/2)\, \sigma_z$, where the $\sigma_i$ are Pauli matrices. Here, $h_x$ sets a fixed scale for off-diagonal coupling between the two basis states of the system, and $V(\tau)$ sets the time-dependent diagonal energy difference between the two levels, with time denoted by $\tau$. Considering the well-known adiabatic rapid passage (ARP) protocol depicted in Fig.~\pref[c]{CDIntro}, in which the diagonal energy difference $V$ is linearly ramped while the off-diagonal coupling $h_x$ is held fixed, a new, effective coupling term  $h_\textrm{CD} \sigma_y$ will appear in the instantaneous reference frame of the particles. This emergent term can lead to transitions between the instantaneous eigenstates, and can result in a breakdown of the ARP protocol.

In extending this example to the many-level scenario shown in Fig.~\pref[b]{CDIntro}, we abandon the Pauli matrices in favor of a formalism that reflects the symmetry of the couplings relevant to a uniform tight-binding lattice. Specifically, the bare Hamiltonian in this case contains off-diagonal coupling between nearest-neighbor states at a scale set by the (real-valued) tunneling energy $t$, along with diagonal site energies that are offset from their nearest neighbors by an amount $V(\tau)$. As in the two-level case, when the energy difference between the neighboring sites is changed dynamically in time, new diabatic terms appear in the instantaneous reference frame. These diabatic terms appear as added contributions to the off-diagonal tunneling that are imaginary, \textit{i.e.}, $\pi/2$ out of phase with respect to the bare, real tunneling. These imaginary tunneling contributions, direct extensions of the emergent $\sigma_y$ term of the two-level case, can be incorporated along with the bare tunneling $t$ into a modified overall tunneling energy $t_\text{CD}$ and tunneling phase $\phi_\text{CD}$. To note, the appearance of such phases underlies the use of shaken-lattice techniques for the dynamical generation of tunneling phases~\cite{Struck-ShakingGauge,Struck-science}. In contrast, the ability to directly engineer tunneling phases in our synthetic lattice platform can allow us to counter-act these dynamical modifications of the tunneling phases, and thus enable faster-than-adiabatic population transport and state transformations.

%
%

\begin{figure}%
\includegraphics{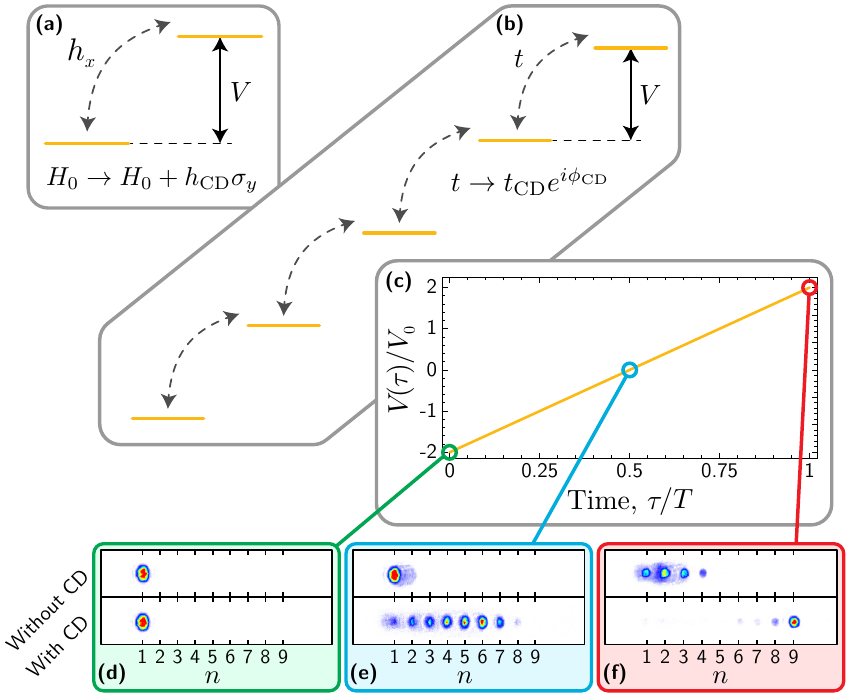}%
\caption{Counterdiabatic driving in multi-level systems.
\pnl{a}~A two-level representation of counterdiabatic driving. The two levels (yellow lines) are subjected to the Hamiltonian $H_0 = h_x \sigma_x +(V\!(\tau)/2)\, \sigma_z$. When the site-energies are swept in time however, diabatic terms proportional to $\sigma_y$ are introduced in the instantaneous reference frame. 
\pnl{b}~An $n$-level representation of counterdiabatic driving. Five sites (yellow lines) in a tight binding model are linked by a constant, real coupling $t$ (dashed black arrows) and all have an energy difference with their neighbors given by the time-dependent function $V(\tau)$. When $V(\tau)$ is changed in time, new effective complex tunneling terms appear.
\pnl{c}~A plot of $V(\tau)/V_0$ versus time $\tau/T$ where $V_0$ is the maximum energy offset and $T$ is the total ramp duration. 
\pnl{d--f}~Averaged absorption images for a nine-site lattice undergoing the site-energy ramp shown in panel {\sffamily (c)} at three different points during the ramp $\tau = \{0, T/2, T\}$ shown with green, blue, and red backgrounds respectively. For this data ${T = \unit[2.5]{ms}}$, ${V_0/h \approx \unit[4]{kHz}}$, and ${t/h \approx \unit[1]{kHz}}$. The experiment is performed both without (top images) and with (bottom images) counterdiabatic driving.
}
\label{CDIntro}
\end{figure}

These diabatic terms (the $\sigma_y$ term in the two-level system and the modified tunneling energy and tunneling phase in the $n$-level system) can be determined analytically. Thus with the proper experimental controls, new terms may be introduced to the Hamiltonian that, when we enter the instantaneous reference frame, cancel with the newly acquired diabatic terms and thus mitigate diabatic transitions between instantaneous eigenstates. This is the idea beyond counterdiabatic (CD) driving, and its ability to overcome the adiabatic limit. We present a lattice example of this effect in Fig.~\pref[d--f]{CDIntro} using ultra cold \textsuperscript{87}Rb in a synthetic lattice of nine coupled momentum states, which will be described in more detail later in this manuscript. The goal of this experiment is to faithfully transfer population across a multi-site lattice (from site ${n=1}$ to ${n=9}$) in a time that is shorter than needed for adiabatic transport. Our nine-site lattice, with an initial tilt of ${V_0/h \approx \unit[4]{kHz}}$ and a final tunneling of ${t/h \approx \unit[1]{kHz}}$, is initialized with all population at its left edge in site ${n=1}$. The site energies and tunneling strengths are then swept according to the Eqs.~\ref{eq:tilttunn}~and~\ref{eq:tiltenergy} appearing later in this manuscript, such that the tilt is inverted in a time ${T = \unit[2.5]{ms}}$.  Averaged absorption images are shown of population in the lattice at three different times along the ramp $\{0, T/2, T\}$ and experiments are performed both without (top images) and with (bottom images) the addition of CD driving. Both experiments start with identical conditions in Fig.~\pref[d]{CDIntro}. By time $T/2$ shown in Fig.~\pref[e]{CDIntro}, however, the CD experiment is already vastly outpacing the adiabatic experiment. Not much transport has occurred without CD driving whereas the population is roughly halfway to the other side of the lattice with CD driving. Figure~\pref[f]{CDIntro} shows the final distributions of the two experiments. Without CD driving, population has only moved a few sites from the starting position. In contrast, CD driving enables the majority of the population to be successfully transported the other side of the lattice. This proof-of-concept experiment shows that CD driving in this case can provide a powerful shortcut to adiabaticity.

\section{Results}

\noindent We present two basic experiments to demonstrate the power of the CD technique and its applicability for state manipulation and preparation in extended, multi-site systems. First, we explore CD-enhanced population transfer across a multi-site lattice. In analogy to ARP protocols~\cite{LZ,PEIK-ARP-Bloch} in two-level systems, we apply a linear energy bias across the sites of a lattice with open boundaries. By inverting this bias ``tilt,'' we attempt to move population from one end of the lattice to the other with high fidelity. Second, we explore how CD methods can aid in preparing the delocalized eigenstates of a multi-site system. Specifically, for a five-site lattice with open boundaries, we prepare and probe its five energy eigenstates.

For both of these experiments, the system is a tight-binding lattice with a Hamiltonian given by
\begin{equation}
H_0 = -\sum_n t_n \left(\create{}\annihilate[n+1]{} +\text{H.c.}\right) + \sum_n V_n \create{}\annihilate{},
\label{eq:cdhzero}
\end{equation}
where $t_n$ is the real, positive tunneling energy associated with a particle transitioning between sites $n$ and $n+1$, $c_n^{(\dagger)}$ is the annihilation (creation) operator at site $n$, and $V_n$ is the energy of the site $n$. Experimentally, the sites of the system are realized by plane-wave momentum states of Bose-condensed $^{87}$Rb atoms, the tunneling elements represent two-photon Bragg transitions driven by applied laser fields, and the site energies are controlled through coordinated detunings of the various Bragg laser fields from their resonance conditions~\cite{Gadway-KSPACE,Meier-AtomOptics}. 

We consider attempting to make an adiabatic change to this Hamiltonian by changing the parameters $V_n$ and $t_n$ as a function of the external control parameter $\lambda(\tau)$, where $\tau$ represents time, such that they become:
\begin{align}
V_n &\rightarrow V_n(\lambda)\\
t_n &\rightarrow t_n(\lambda).
\end{align}
The parameter $\lambda$ encapsulates the time-dependence of the Hamiltonian but could in principle be a function of an external field or any other variable. For general time-dependence of the system parameters, the instantaneous eigenstates of the system can become coupled by new, diabatic terms in the Hamiltonian~\cite{driespnas, demirplak2003,demirplak2005, berry2009}. For this particular Hamiltonian, the dominant diabatic terms, 
\begin{align}
A=i\sum_n \alpha_n \left(c^\dagger_{n}c_{n+1}-c^\dagger_{n+1} c_{n} \right),   
\end{align}
take the form of a local current, where $\alpha_n$ can be found by solving a closed set of equations. This system of equations, derived in Sec.\,\ref{app:deriv1} and based on the supplementary information of Ref.~\cite{driespnas}, is given by
\begin{align}
\begin{split}
-t_n \partial_\lambda \left( V_{n+1} - V_n \right) = &+\alpha_n \left( V_{n+1} - V_n \right)^2 \\ &+\alpha_n \left( t_{n+1}^2 + 4 t_n^2 + t_{n-1}^2 \right)\\ &- 3 t_n \left( t_{n+1} \alpha_{n+1} + t_{n-1}\alpha_{n-1} \right),
\label{eq:cdeqns}
\end{split}
\end{align}
where every term is a function of $\lambda$ and so the explicit $(\lambda)$ notation has been suppressed. For simplicity, this set of equations includes only the spatial dependence of the tunneling terms and not the time-dependence. For the experiments performed herein, the diabatic effects come almost entirely from the time-dependence of the site energies and not the tunneling terms, which only negligibly change the $\alpha_n$ values. For a full derivation, including all time-dependencies, see Sec.\,\ref{app:deriv2}.

In our experimental implementation of counterdiabatic driving, we directly incorporate these diabatic terms into our engineered Hamiltonian through control of the tunneling amplitudes and tunneling phases as: 
\begin{align}
\begin{split}
H_\text{CD}(\lambda) = &-\sum_n \left(t_{n,\text{CD}}(\lambda) e^{-i \phi_{n,\text{CD}}(\lambda)} \create{}\annihilate[n+1]{} +\text{H.c.}\right)\\ &+ \sum_n V_n(\lambda) \create{}\annihilate{},
\label{eq:cdhcd}
\end{split}
\end{align}
where the new terms $t_{n,\text{CD}}(\lambda)$ and $\phi_{n,\text{CD}}(\lambda)$ are given by
\begin{align}
t_{n}(\lambda) &\rightarrow t_{n,\text{CD}}(\lambda) = \sqrt{t_n(\lambda)^2 + \left(\alpha_n(\lambda) \partial_\tau \lambda(\tau) \right)^2},\label{eq:cdcorrectedtunn}\\
\phi_n &\rightarrow \phi_{n,\text{CD}}(\lambda) = \arctan\!\left( \frac{t_n(\lambda)}{\alpha_n(\lambda) \partial_\tau \lambda(\tau)} \right).
\label{eq:cdcorrectedphase}
\end{align}

While the incorporation of the additional control variable $\lambda$ may seem like an unnecessary extra step, the solution in terms of $\lambda$ instead of $\tau$ allows one to solve Eq.~\ref{eq:cdeqns} without prior knowledge of the explicit functional form of the time-dependence. As long as it is known how $t_n$ and $V_n$ relate to $\lambda$, the functional form of $\lambda$ can be changed without re-solving for the $\alpha_n(\lambda)$. The site- or $n$-dependence of these diabatic corrections, seen in  Eqs.~\ref{eq:cdcorrectedtunn}~and~\ref{eq:cdcorrectedphase}, make this direct form of CD driving only well matched to certain experimental implementations that allow for local control of state-to-state coupling (tunneling) strengths and phases, such as is found in ``synthetic lattices.'' For implementation in other platforms, there exists a remapping of the diabatic corrections through a gauge transformation that allows them to be applied solely to the site energies. Given our ability to directly control the tunneling strengths and phases in our system, we implement the direct, ungauged version of the CD protocol.

\subsection{Transfer across the lattice}

\begin{figure*}
\includegraphics{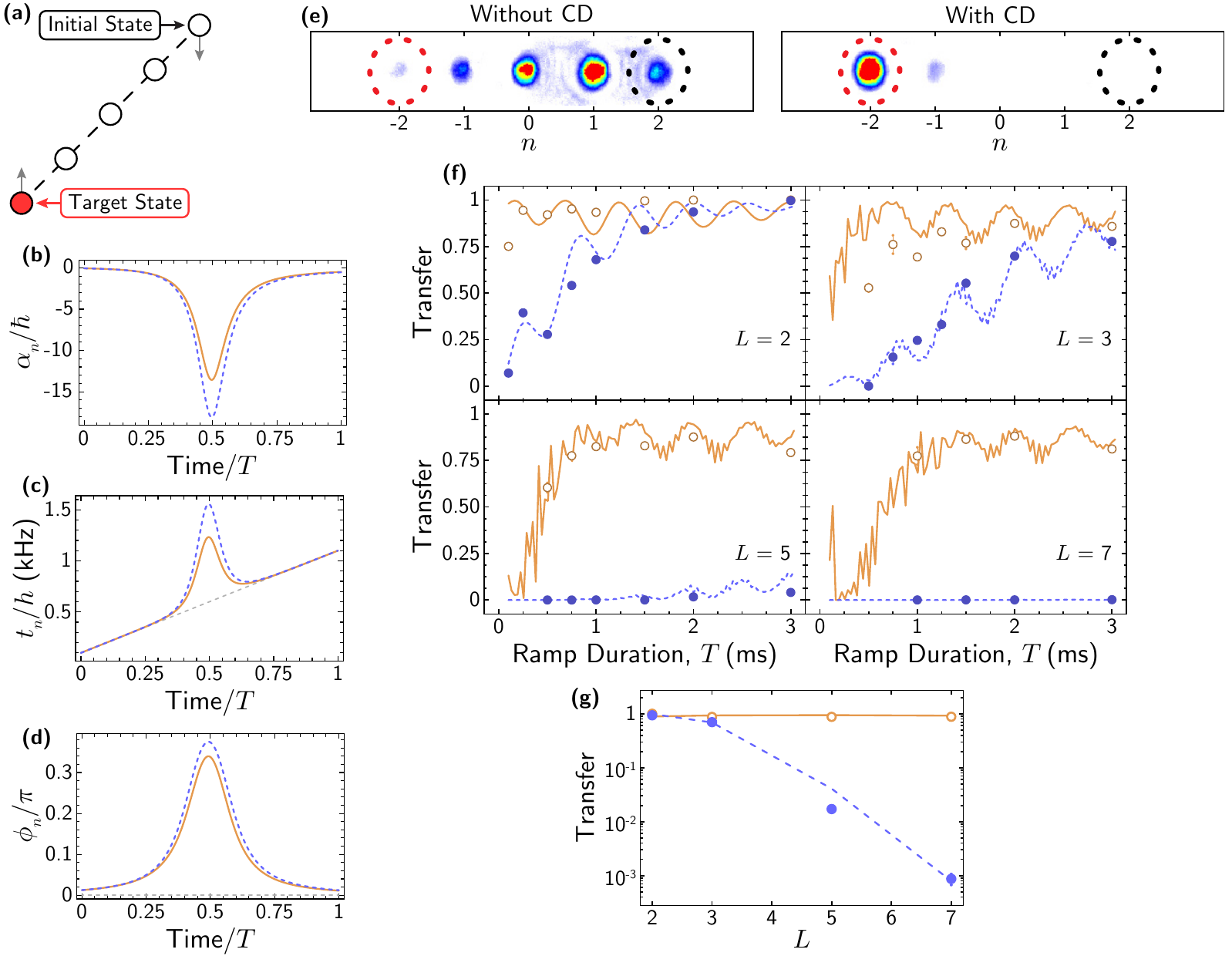}
\caption[Enhancement of population transfer by CD driving]{Enhancement of population transfer by CD driving.
\pnl{a}~Cartoon depiction of the initial, highest-energy state and the target state for data shown in {\sffamily(e)}.
\pnl{b--d}~The $\alpha_n$, $t_{n,\text{CD}}$, and $\phi_{n,\text{CD}}$ for a five site lattice with $n \in \{-2,-1,0,1,2\}$ and $T=\unit[2]{ms}$. The blue dotted line represents the terms for the transitions $-1\rightarrow 0$ and $0 \rightarrow 1$ and the gold solid line is for $-2\rightarrow -1$ and $1 \rightarrow 2$. The gray dashed lines in {\sffamily(c)} and {\sffamily(d)} show $t_n$ and $\phi_n$ without CD corrections.
\pnl{e}~Averaged (over eight measurements) absorption images for a $T=\unit[2]{ms}$ ramp in an $L=5$ site lattice with $V_0/h \approx \unit[4]{kHz}$ and $t_n/h\approx \unit[1]{kHz}$. The procedure is performed both without CD driving (left) and with (right). The target state is circled with the red dotted line on the left and the starting state is on the far right circled with the black dotted line.
\pnl{f}~The population transferred from the initial state to the target state at the end of the ramp versus the total ramp duration, $T$. Open orange circles and solid orange lines represent data and simulation when implementing CD driving during the ramp. Filled blue circles and dashed blue lines represent data and simulation without implementing CD driving during the ramp. We perform the protocol with $L=2$, $L=3$, $L=5$, and $L=7$.
\pnl{g}~Log of population fraction transferred to the target state as a function of system size $L$ for a $T=\unit[2]{ms}$ ramp. Open orange circles and solid orange line are data and simulation when using CD driving during the ramp. Filled blue circles and dashed blue line are data and simulation without CD driving during the ramp.
All error bars indicate one standard error of the mean. The data shown here were taken with $t/h\approx\unit[1]{kHz}$ and the simulations use an average mean-field interaction of $U/h=\unit[1.25]{kHz}$.
}
\label{CDtransfer}
\end{figure*}

\noindent Our first demonstration of how CD driving can provide a shortcut to adiabaticity is in the context of a multi-site ARP. We begin with all the population in the highest excited state of a lattice with a variable total number of sites $L$ as depicted in Fig.~\pref[a]{CDtransfer}. In the absence of CD driving, we linearly ramp the lattice from a positive tilt to a negative tilt. The highest energy mode of the system goes from initially populating the right-most site at the start of the ramp, to having nearly full overlap with the left-most site at the ramp's end. The parameter ramp we implement, which has a total duration $T$, has the exact form
\begin{equation}
\lambda(\tau) = 1 - \frac{\tau}{T}.
\label{eq:cdtiltlambda}
\end{equation}
This ramp is incorporated into $t_n$ and $V_n$ as
\begin{align}
t_n(\lambda) &= t (1.1-\lambda) = t \left(0.1+\frac{\tau}{T}\right)\label{eq:tilttunn}\\
V_n(\lambda) &= n V_0 2(\lambda-1/2) = n V_0 \left(1 - \frac{2\tau}{T}\right),\label{eq:tiltenergy}
\end{align}
where $t$ is the characteristic tunneling scale of the lattice at the end of the ramp and $V_0$ is the initial site energy slope. With the large energy offset and small tunneling, the eigenstates of the initial Hamiltonian are essentially site-localized. The tunneling is independent of $n$, meaning that each link always has the same tunneling strength (in the adiabatic protocol) whereas the site energy is linear with $n$ such that the lattice is tilted with a slope $V_0$ at $\tau=0$ and $-V_0$ when $\tau=T$. The correction factors $\alpha_n$ as calculated from Eq.~\ref{eq:cdeqns} are plotted along with the modified tunneling strength and phase (from Eqs.~\ref{eq:cdcorrectedtunn}~and~\ref{eq:cdcorrectedphase}) in Figs.~\pref[b--d]{CDtransfer}, illustrating the kinds of corrections that can be generated by the CD approach. 

The calculated $\alpha_n$ values shown in Figs.~\pref[b--d]{CDtransfer} are for a five-site lattice undergoing this inversion in a time $T=\unit[2]{ms}$ with an initial offset $V_0/h \approx \unit[4]{kHz}$ and a characteristic tunneling $t/h\approx \unit[1]{kHz}$. The $\alpha_n$ terms are then used to calculate the CD tunneling strength $t_{n,\text{CD}}(\lambda)$ and phase $\phi_{n,\text{CD}}(\lambda)$. The diabatic terms $\alpha_n$ are symmetric about the center of the lattice in this case, such that $\alpha_{-1} = \alpha_{0}$ (blue dashed line) and $\alpha_{-2}=\alpha_{1}$ (solid gold line) as shown in Fig.~\pref[b]{CDtransfer}. This symmetry propagates onto the tunneling strengths and phases as well, as shown in matching style in Figs.~\pref[c,\,d]{CDtransfer}. The uncorrected tunneling strength and phase are also shown in these figures as gray dashed lines for comparison.

If performed slowly enough so as to respect adiabaticity conditions, this lattice tilt inversion would move population from the right end of the lattice to the left end, maintaining population of the highest energy eigenstate throughout the evolution. However, if the lattice tilt inversion is performed too rapidly, it can induce non-adiabatic Landau--Zener transitions between the instantaneous eigenstates of the system. For a two-site lattice, corresponding to the textbook ARP, such a breakdown in adiabaticity would result in some population remaining at the initial site. For the many-site system, this breakdown in adiabaticity more generally results in atomic population not making it all the way across the lattice, as a result of undergoing one or more diabatic transitions between the system eigenstates. Figure~\pref[e]{CDtransfer} shows this effect in the averaged absorption image data for $L=5$, $t/h \approx \unit[1]{kHz}$, $V_0/h \approx \unit[4]{kHz}$, and $T=\unit[2]{ms}$. Without CD driving (left image in the panel), we detect only a small fraction of atoms in the left target site due to a breakdown in adiabaticity during the lattice inversion. In fact, most atoms remain within one or two sites of the initial position. In contrast, the application of CD driving (right image in the panel) results in nearly all of the atoms transferring to the left-most target site after starting in the right-most site.

We perform this lattice inversion experiment as a function of both ramp duration $T$ and system size $L$ and present the results in Fig.~\pref[f]{CDtransfer}. To note, we keep $t/h \approx \unit[1]{kHz}$ and $V_0/h \approx \unit[4]{kHz}$ fixed as we vary the ramp duration $T$ and system size $L$. The specific $t$ and $V_0$ we work with here are chosen to optimize the larger system sizes and are not ideal for the shorter system sizes (\textit{i.e.}\ protocols may be performed much faster for 2- and 3-site lattices than is shown here). On the vertical axis we plot the transfer fraction (the fractional amount of atoms which ended up in the target state) against the ramp duration $T$ on the horizontal axis. Focusing on the $L=2$ case presented in the upper left, we find that implementing CD driving results in a substantial increase in the transferred fraction for shorter ramps as reflected in the data (open orange circles) and simulation (solid orange line) over the case without CD driving (solid blue dots and dashed blue theory line). Without CD driving, we are eventually able to reach adiabaticity in this small system for ramp durations $T > \unit[2]{ms}$. Indeed, in this limit and for just two sites, the diabatic terms become negligible and the CD and non-CD parameter ramps are essentially identical. 

We also perform this lattice transfer experiment as a function of the system size $L$ working with sizes ${L\,=\,\{2,3,5,7\}}$. We observe, unsurprisingly, that without CD driving it becomes increasingly difficult to maintain adiabaticity as the system size grows. This is reflected in Fig.~\pref[f]{CDtransfer} by the continued reduction of the transfer fraction for the blue curves/points as $L$ is increased, even out to the longest times explored. With the application of CD driving, however, we are able to transfer ${>80\%}$ of population to the target state for all four system sizes on the timescale of just 1--2 ms. For the shortest ramps explored, we observe one practical limitation of the CD driving as applied to our ``momentum-state synthetic lattices,'' which results in a decreased transfer efficiency. Specifically, the CD corrections become quite large for short ramps, and the operational limit of our effective lattice Hamiltonian breaks down as different momentum states become coupled in an off-resonant fashion.

The dependence of the non-CD and CD transfer efficiency on the system size $L$ is summarized in Fig.~\pref[g]{CDtransfer}. We show the population transferred to the target state as a function of system size for a ${T=\unit[2]{ms}}$ ramp duration for all four system sizes on a semi-log scale. We observe that the CD driving (open orange circles) results in very near to perfect transfer for all lattice sizes, whereas the case without CD driving (solid blue dots) results in a transfer which decreases very quickly with system size, eventually reaching the 10\textsuperscript{-3} level for ${L=7}$. This plot emphasizes the enormous benefits of using CD driving when performing a non-trivial operation in a system containing an adiabatic limit that scales poorly with system size (\textit{i.e.}, in systems with no adiabatic support in the thermodynamic limit). Such state transfers may be of practical benefit---for example, in this system of coupled momentum states, the measured efficiency of 88\% for transferring across 7 synthetic lattice sites in \unit[2]{ms} relates to an efficiency of over 99\% per imparted recoil momentum ($\hbar k$, with $k$ the lattice wave vector).

\subsection{Preparing and probing eigenstates of few-site lattices}

\begin{figure*}
\centering
	\includegraphics{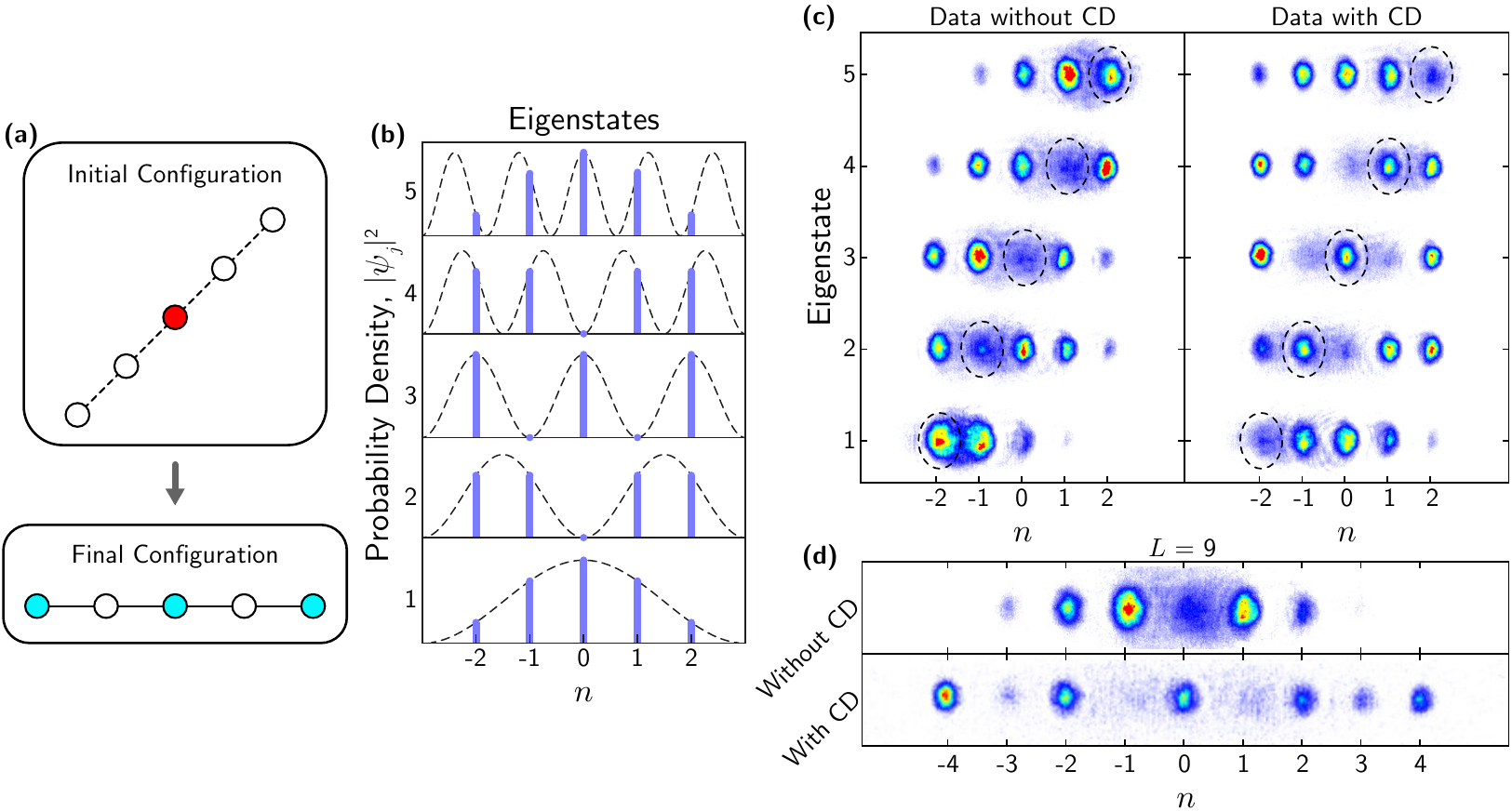}
	\caption[Preparing the eigenstates of an $L$-site lattice with counterdiabatic driving]{
		Preparing the eigenstates of an $L$-site lattice with counterdiabatic driving.
		\pnl{a}~Depiction of the ramp protocol for eigenstate preparation. We start in a single site of the initial configuration and the choice of this site dictates which eigenstate is prepared. By smoothly turning off a large gradient, eigenstates of the flat, uniform lattice are prepared. We show as an example how by starting at the central site of the tilted lattice can allow for preparation of the middle-energy eigenstate of a five-site lattice.
		\pnl{b}~The probability density for the first five eigenstates of a five-site lattice. The dashed lines indicate the probability for a continuous ``infinite square well'' potential, whereas the blue bars show the discretized amplitudes relating to a five-site lattice with open boundaries.
		\pnl{c,\,left}~Absorption images for the attempted preparation of all five eigenstates without CD driving. We compare the theoretical probability distribution and our adiabatically prepared probability distribution by calculating the ``efficiency'' $F_j^\text{adi} =$ \{0.722(24), 0.555(11), 0.282(15), 0.619(9), 0.808(7)\} for each of the five eigenstates.
		\pnl{c,\,right}~Absorption images for the preparation of all five eigenstates with CD driving during the ramp. For these five states, the associated preparation ``efficiencies'' are $F_j^\text{cd} =$ \{0.983(5), 0.861(5), 0.865(3), 0.911(8), 0.987(2)\}. Data shown in {\sffamily(c)} were taken with a ramp time of \unit[1]{ms} and a final tunneling of $t/h = \unit[950]{Hz}$. The dashed circles shown for each image in {\sffamily(c)} relate to the initial locations of the atomic condensate. Typically, a small residual cloud of thermal atoms remain at this initial location relating to zero momentum.
		\pnl{d}~Absorption images showing the attempt to prepare the middle-energy state for a nine-site lattice taken with a ramp time of \unit[1]{ms} and a final tunneling of $t/h = \unit[1]{kHz}$. The upper panel is without CD driving and the lower panel is with CD driving.
		}
	\label{CDEigenstates}
\end{figure*}

\noindent In addition to performing robust state transformations for practical applications, CD driving also promises to greatly improve the ability to prepare specific wavefunctions that may be of interest to the study of localization phenomena in synthetic lattices. For example, the physics of topological boundary states~\cite{Meier-SSH,Meier-TAI} could be better probed by engineering tailored atomic wave functions that are well matched to boundary modes. In addition, the ability to engineer individual energy eigenstates of certain types of quasiperiodic lattice models could allow for the direct exploration of localization phenomena at a mobility edge~\cite{Ganeshan-Pixley-GAA,Biddle-T1T2,Biddle-ExpT,An-MobEdge,Luschen-Bloch-SPME}. Finally, for synthetic lattices of momentum states, as well as for techniques allowing for engineered spin-orbit coupling in atomic gases~\cite{Galitski-Spielman}, the physical separation in space of atomic wavepackets with distinct momentum can present a practical limitation on the timescales over which coherent nonequilibrium dynamics can be explored. By preparing ``dressed'' eigenstates (dressed states of the atoms and driving fields) that involve superpositions of multiple different plane-wave momentum states, one can judiciously explore the physics of such systems while circumventing the issue of spatial wavepacket separation.

Here, for demonstration purposes, we consider populating the dressed eigenstates of a simple five-site synthetic lattice by the methods of adiabatic preparation and CD driving. Our starting configuration is a 5-site lattice with the same initial positive site-energy slope as used in the previous experiment. Here, however, instead of continuing to invert the site energy tilt from positive slope to negative slope, we terminate this procedure at the time when the sites of the lattice all have equal energy values, as illustrated in Fig.~\pref[a]{CDEigenstates}. The form of the parameter ramp function $\lambda(\tau)$, as well as the ramps for the tunneling parameters, are identical to the previous case (\textit{i.e.}, Eqs.~\ref{eq:cdtiltlambda}~and~\ref{eq:tilttunn}), but now we ramp the site energies as
\begin{align}
V_n(\lambda) = n V_0 \lambda = n V_0 \left( 1 - \frac{\tau}{T} \right),
\end{align}
where $n$ is the site index, $V_0$ is initial slope of the tilted lattice, $T$ is the ramp duration, and $\tau$ is the time variable. The diabatic terms and corrections to the tunneling amplitudes and phases for this experiment look identical to those appearing in the first half of the lattice inversion experiment shown in Figs.~\pref[b--d]{CDtransfer}. This new ramp results in an initial positive slope of $V_0$ at $\tau=0$, but instead of fully inverting, the site energies become equal at $\tau=T$, resulting in a flat and uniform lattice.

The Bloch-like eigenstates of this uniform lattice are essentially just the lowest five eigenstates of a ``particle in a box'' or infinite square well, however with the wavefunction amplitude only appearing at the five discrete positions corresponding to the sites of the lattice, shown in Fig.~\pref[b]{CDEigenstates}. While the higher energy eigenstates should indeed feature more rapid variation of the phase, as is related to the appearance of nodes in the wave functions in the continuous problem, there is a symmetry of the low- and high-energy eigenstates if one looks only at their probability densities. Specifically, as is depicted by the heights of the blue bars in Fig.~\pref[b]{CDEigenstates}, the lowest and highest energy eigenstates (eigenstates 1 and 5) should have equivalent probability density distributions, as should the eigenstates with the second-lowest energy (state 2) and the second-highest energy (state 4).

In the previous experiment (state-transfer by lattice tilt inversion), we specifically loaded the highest-energy eigenstate to start. This was accomplished by starting with all population initialized to our atomic condensate at zero momentum, and then only turning on the appropriate Bragg laser fields that would address transitions to atomic momentum states displaced in one direction (\textit{e.g.}, to states with two photon momenta to the left, four photon momenta to the left, etc.). In general, we have full control over which transitions between momentum states are ``turned on'' through laser-addressing, Physically, we always begin in the zero momentum state, which is denoted by the dashed circles in Fig.~\pref[c]{CDEigenstates} (and about which one can typically see a small cloud of residual thermal atoms). However, the mapping between physical ``momentum state'' and ``synthetic lattice site'' can be modified to allow for the initialization at different sites of the five-site tilted lattice, which in turn allows for the initialization of different eigenstates in the tilted lattice configuration (as depicted in Fig.~\pref[a]{CDEigenstates}). In this way we can prepare all five eigenstates $\ket{\psi_j}$ of the uniform open-boundary lattice, first by preparation with an energy bias applied and then by a ramp of the tilt to zero site-to-site bias.

We first attempt this experiment without CD driving, as shown in Fig.~\pref[c,\,left]{CDEigenstates} with the parameters $V_0/h \approx \unit[4]{kHz}$, $t/h \approx \unit[0.95]{kHz}$, and $T=\unit[1]{ms}$. It is qualitatively clear from the absorption images that the procedure without CD driving failed to prepare any of the eigenstates, as the probability densities do not match those expected for the eigenstates, as shown in Fig.~\pref[b]{CDEigenstates}. More to the point, the atomic distribution did not spread out much from the initial zero-momentum condensate that was first populated (dashed circles). Quantitatively, we compare the theoretical probability distribution and our adiabatically prepared probability distribution by calculating the ``efficiency'' of state preparation. We calculate the efficiency in terms of the normalized number of atoms detected at each lattice site $n$ and for each state $j$ which we define as $P_n^j$. The ``efficiency'' $F$ is calculated from these probability distributions by 
\begin{align}
F_j = \left(\sum_n \sqrt{P_n^{j,\,\text{theory}}P_n^{j,\,\text{expt}}}\right)^2
\end{align} 
in terms of the experimental and theoretical probability distributions. In the adiabatic case ${F_j^\text{adi}=}$~\{0.722(24), 0.555(11), 0.282(15), 0.619(9), 0.808(7)\} for each of the five eigenstates. To note, the adiabatic approach works most efficiently for the highest and lowest energy states, which have the fewest neighboring (in energy) eigenstates to which they could undergo diabatic transitions.

We observe that a marked improvement is obtained by applying CD driving to this state preparation protocol, keeping all other parameters the same. We are able to prepare probability distributions which agree well with the desired distributions, with ${F_j^\text{cd}=}$~\{0.983(5), 0.861(5), 0.865(3), 0.911(8), 0.987(2)\} for eigenstates 1--5, with corresponding experimental absorption images shown in Fig.~\pref[c,\,right]{CDEigenstates}. Thus the average efficiency for preparing states via the adiabatic method is 0.60(5), with an improvement to an average efficiency of 0.92(3) via the CD method. Just as for the state transfer across the lattice, this preparation of eigenstates delocalized across the sites can also be extended to larger system sizes. Fig.~\pref[d]{CDEigenstates} shows (via absorption image) the preparation of the middle-energy eigenstate for an $L=9$ site lattice, which should have the same alternating probability density structure as the middle-energy in Fig.~\pref[b]{CDEigenstates}. The upper panel shows the experiment as performed without CD driving and relates to an efficiency of ${\sim 50\%}$. The lower panel, with CD driving, matches the expected distribution much more closely with an efficiency of ${\sim 80\%}$.

These measurements show that we are able to accurately prepare a state which has the same probability density distribution as the desired system eigenstates. However, we have not yet characterized the phase structure of the created wave functions, which is of equal importance. One simple and generic way to test whether we have created the appropriate phase structure for our states (in addition to the probability density distribution) is to simply \textit{leave on} the base Hamiltonian of a flat lattice with uniform tunnel-coupling, and observe whether there is continued evolution of the probability density distribution. If the phase structure is matched to that of the desired \textit{eigenstate}, then by definition we should see no further dynamics of the population density under that Hamiltonian's evolution. However, any errors in the populations or relative phases between sites can drive currents in the lattice.

\begin{figure}
	\includegraphics{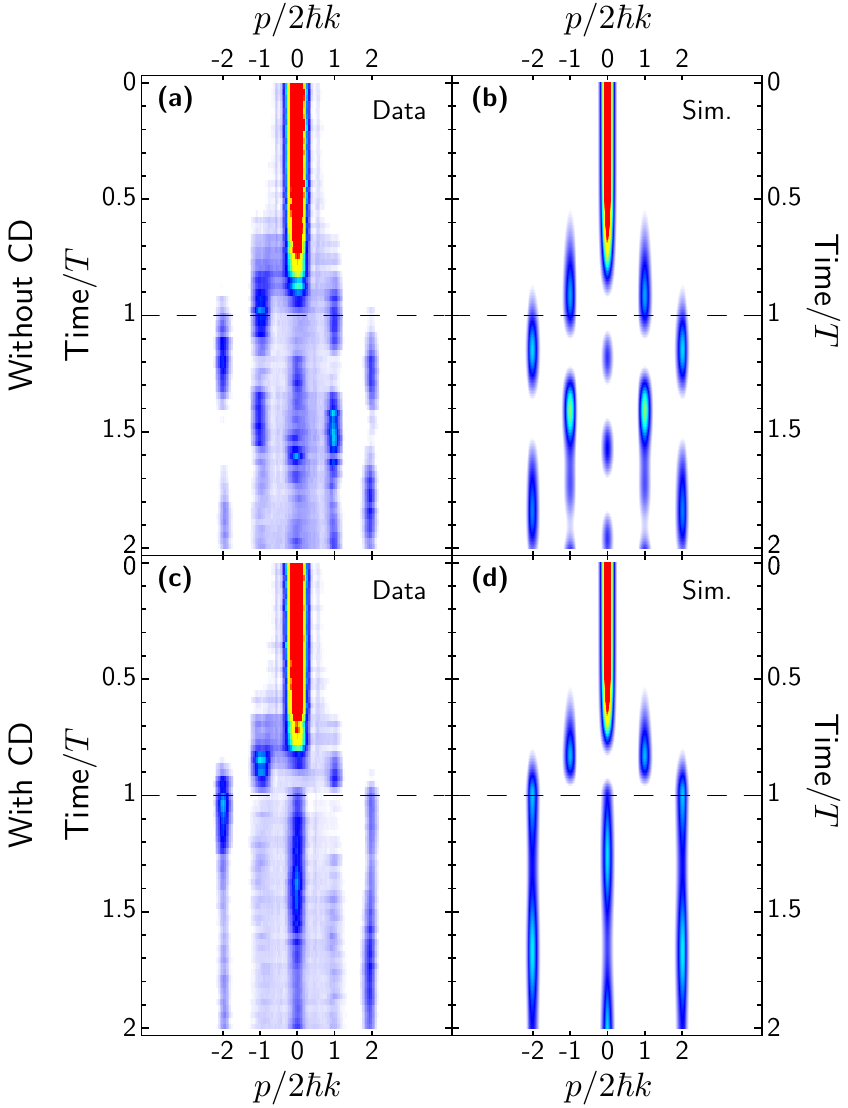}
	\caption{
		Testing the phase structure of the counterdiabatically prepared eigenstates.
		Observations of any dynamics after the ramp when preparing the middle-energy state of a five-site lattice for ramps without CD driving [data in \pnl{a} and simulation in \pnl{b}] and for ramps with CD driving [data in \pnl{c} and simulation in \pnl{d}]. The dashed lines in all four panels indicate the end of the ramp. Data shown here were taken with a ramp time $T=\unit[0.75]{ms}$ and a final tunneling $t/h=\unit[900]{Hz}$.}
	\label{CDphase}
\end{figure}

Figure~\pref{CDphase} summarizes our tests of the phase structure of the prepared states. We perform the same protocol used to prepare the eigenstates of the 5-site lattice but with parameters $V_0/h\approx \unit[4]{kHz}$, $t/h \approx \unit[900]{Hz}$, and $T = \unit[0.75]{ms}$. We attempt to prepare eigenstate 3 of the 5-site lattice as shown in Fig.~\pref{CDEigenstates} both with and without CD driving with the added step of keeping the Hamiltonian on and static after the ramp period for an additional time $T$. Without CD driving, we observe continued dynamics after the end of the ramp as shown in Fig.~\pref[a]{CDphase}. This indicates that we have failed to prepare the eigenstate, which is no surprise since even the probability density of the prepared state did not match the eigenstate at the end of the ramp (black dashed line). In addition, the data agree with a numerical simulation of the ``ramp and hold'' dynamics, as shown in Fig.~\pref[b]{CDphase}. In contrast, when we attempt to prepare the eigenstate with the aid of CD driving, we find there is very little dynamics during the hold period after the ramp of the lattice tilt. This is shown both for the data and the simulations in Figs.~\pref[c,\,d]{CDphase}. This signifies that we have nearly correctly prepared the desired eigenstate in both the probability amplitude and phase structure. We note that the small amount of residual time dependence is actually expected, based on our still slightly imperfect preparation of the eigenstate. This is in part due to the role of atomic interactions~\cite{An-Inter}, which are not incorporated into the design of the employed CD protocol.

\section{Discussion}

\noindent We have demonstrated that CD driving can lead to drastic improvements over simpler adiabatic schemes in relation to population transfer and state preparation in multi-level, lattice-like systems. We showed that the CD approach can lead to significant improvement in the preparation of system eigenstates, which will aid in exploration of tight-binding models with novel eigenstates, as well as the engineering of laser-``dressed'' atomic momentum states.
These experiments represent the first implementation of a truly non-trivial shortcut to adiabaticity protocol, \textit{i.e.}, in a system beyond just two or three states that does not possess scaling symmetry. As such they prove the power of this technique to transform states in systems with small gaps or poor adiabatic scaling. Indeed, this regime of many-level systems is where the use of these shortcuts yields the most dramatic improvements over adiabatic protocols. The specific fast, robust, and efficient manipulations we have explored promise to provide practical advantages for applications such as atom interferometry, and to enable new explorations of exotic phenomena related to topology and disorder in synthetic lattices. By incorporating controlled loss in our synthetic lattice~\cite{Lapp_2019,YanLoss}, such approaches may also allow for unique explorations into faster-than-adiabatic approaches in non-Hermitian systems~\cite{Shortcut-NonHermitian}.

In this work, we have largely forgone any discussion of the role of interactions in our atomic system~\cite{An-Inter}. While the effects of mean-field level interactions were included in the simulations of Fig.~\pref{CDtransfer} at the relevant experimental levels (see App.\,\ref{app:GPE}), they did not result in significant deviations from the non-interacting scenario. This is because we have restricted the present exploration to the regime in which the lattice bandwidth (four times the tunneling energy) was typically much larger than the mean-field interaction energy. One straightforward extension of this work would be to harness the natural interactions in our system of Bose-condensed atoms and to investigate optimal CD driving protocols in the presence of strong nonlinear interactions~\cite{saberi2014,campbell2015,bukov2019,Hatomura2017,Hatomura2018}.

The combination of native interactions in our system and the ability to explore CD driving may also enable future explorations of topological defects in interacting analog systems. Specifically, there exists a formal connection between the diabatic errors that are generated by the change of a Hamiltonian and the Berry curvature of the underlying Hamiltonian parameter space~\cite{Gritsev6457}. This connection has enabled previous measurements of Chern numbers for non-interacting Abelian~\cite{Gritsev6457,schroer2014} and non-Abelian~\cite{PhysRevLett.117.015301,Sugawa1429} analog systems based on superconducting qubits and cold atoms, as well as explorations into how interactions enrich this problem for an interacting pair of qubits~\cite{Roushan2014}. The interactions in our system could allow for natural extensions of these results, enabling investigations of interaction-stretched monopoles~\cite{Ho-monopole} and emergent Yang monopoles~\cite{Zhou-YangMonopole} in a cold atom simulator.

\section{Acknowledgements}
\noindent This material is based upon work (E.J.M.\ and B.G.) supported by the Air Force Office of Scientific Research under Grant No.\ FA9550-18-1-0082. The authors thank A.~Polkovnikov for insightful discussions. D.S. acknowledges support of the Research Foundation Flanders (FWO).

\bibliographystyle{apsrev4-1}
\bibliography{CDarXiv}

\clearpage


\appendix

\section{Derivation of counter-diabatic equations}

\subsection{Time-dependent energy and spatial-dependent tunneling}
\label{app:deriv1}
\noindent Here we derive the optimal local counter-diabatic gauge for transitionless driving of a general free fermion problem. We will be focus on the Hamiltonians of the form 
\begin{equation}
H=-\sum_j t_j \left( c^\dagger_{j} c_{j+1} +h.c. \right)+ \sum_j V_j(\lambda)  c^\dagger_{j} c_j,
\end{equation}
where $c^\dagger_j$ creates a fermion on site $j$ and $c_j$ annihilates the fermion. The approximate adiabatic gauge potential defining CD driving should minimize the following action (see Ref.~\cite{driespnas}):
\begin{equation}
\mathcal{S}\left(\mathcal A_\lambda\right)=\frac{{\rm Tr}\left[G_\lambda^2\right]}{{\rm Tr} \mathbb{I}},
\label{eq:supl_CDaction}
\end{equation}
where 
\begin{equation}
G_\lambda=\partial_\lambda H+i\left[\mathcal A_\lambda,H \right] .
\end{equation}
For quadratic problems the adiabatic gauge potential is also quadratic. Because it is also imaginary it has to be expressed in the form:
\begin{equation}
\mathcal A_\lambda=i \sum_{j,k} \alpha_{j,k}\left( c^\dagger_{k} c_j -h.c.\right),
\end{equation}
where $\alpha_{j,k}=-\alpha_{k,j}$ and all elements are real. Here, we are not concerned with finding the exact adiabatic gauge potentials but rather in their best local approximations, as this drive can implemented in the experiment by appropriately tuning the amplitude and phase of the lasers. Thus, as in the main text, we are restricting $\mathcal A_\lambda$ to the following form
\begin{equation}
\mathcal A_\lambda=i \sum_j \alpha_j\left( c^\dagger_{j} c_{j+1} -h.c.\right),
\end{equation}
where the $\alpha_j$'s remain to be determined by minimizing~\eqref{eq:supl_CDaction}. It is straightforward to check that
\begin{align}
G_\lambda= &\sum_j \left(\partial_\lambda V_j - 2(t_j\alpha_j-t_{j-1}\alpha_{j-1})\right) c_j^\dagger c_j \nonumber\\
&+\sum_j  \alpha_j (V_{j}-V_{j+1}) (c_{j+1}^\dagger c_j+c_j^\dagger c_{j+1}) \nonumber \\
& +\sum_j (t_j\alpha_{j+1}-\alpha_{j}t_{j+1})(c_{j+2}^\dagger c_{j}+c_{j}^\dagger c_{j+2}).
\label{eq:supl_Glambda}
\end{align}
Up to the terms independent of $\mathcal A_\lambda$, it follows from, \textit{e.g.}, Wick's theorem that the action~\eqref{eq:supl_CDaction} is simply proportional to the sum of squares of the individual contributions in the expression above:
\begin{align}
\mathcal S(\mathcal A_\lambda)=&{\ \rm const}+{\frac{1}{4}} \sum_j \left(\partial_\lambda V_j - 2(t_j\alpha_j-t_{j-1}\alpha_{j-1})\right)^2\nonumber \\
&+{\frac{1}{2}} \sum_j \left[
\left(t_j\alpha_{j+1}-\alpha_{j}t_{j+1}\right)^2 +(V_{j+1}-V_j)^2\alpha_j^2\right] .
\end{align}
Minimizing the action with respect to $\alpha_j$ yields the following set of linear equations
\begin{align}
&-3(t_jt_{j+1})\alpha_{j+1}+\left(t_{j-1}^2+4t_j^2+t_{j+1}^2 \right)\alpha_j-3(t_jt_{j-1})\alpha_{j-1} \nonumber \\
&+(V_{j+1}-V_j)^2 \alpha_j=-t_j(\partial_\lambda V_{j+1}-\partial_\lambda V_j).
\label{eq:supl_vargauge_discrete}
\end{align}
In some specific cases these equations can be solved analytically but in general one has to resort to numerical methods.

\subsection{Time-dependent energy and spatiotemporal-dependent tunneling}
\label{app:deriv2}

\noindent One can go through exactly the same exercise as in the previous section, but for a time-dependent tunneling rather than a time dependent potential. Because of the linearity of a problem, a general counter-diabatic drive for time-dependent tunneling and potential will simply be the sum of two independent drives. 
Consider, 
\begin{equation}
H=-\sum_j t_j(\lambda) \left( c^\dagger_{j} c_{j+1} +h.c. \right)+ \sum_j V_j  c^\dagger_{j} c_j,
\end{equation}
In order to compute the action~\eqref{eq:supl_CDaction}, we can recycle most of our previous results. We simply need to replace a few terms in expression~\eqref{eq:supl_Glambda}, resulting in 
\begin{align}
G_\lambda= -2&\sum_j \left(t_j\alpha_j-t_{j-1}\alpha_{j-1}\right) c_j^\dagger c_j \nonumber\\
&+\sum_j  \left[ -\partial_\lambda t_j+ \alpha_j (V_{j}-V_{j+1})\right] (c_{j+1}^\dagger c_j+c_j^\dagger c_{j+1}) \nonumber \\
& +\sum_j (t_j\alpha_{j+1}-\alpha_{j}t_{j+1})(c_{j+2}^\dagger c_{j}+c_{j}^\dagger c_{j+2}).
\label{eq:supl_Glambda_t}
\end{align}
Such that the action reads
\begin{align}
\mathcal S(\mathcal A_\lambda)=&{\ \rm const}+ \sum_j \left( t_j\alpha_j-t_{j-1}\alpha_{j-1} \right)^2\nonumber \\
&+{\frac{1}{2}} \sum_j 
\left(t_j\alpha_{j+1}-\alpha_{j}t_{j+1}\right)^2   \nonumber \\
+&{\frac{1}{2}} \sum_j  \left(\partial_\lambda t_j+(V_{j+1}-V_j)\alpha_j\right)^2 .
\end{align}
Minimizing this action with respect to $\alpha_j$ yields the following set of linear equations
\begin{align}
&-3(t_jt_{j+1})\alpha_{j+1}+\left(t_{j-1}^2+4t_j^2+t_{j+1}^2 \right)\alpha_j-3(t_jt_{j-1})\alpha_{j-1} \nonumber \\
&+(V_{j+1}-V_j)^2 \alpha_j=-\partial_\lambda t_j( V_{j+1}- V_j).
\label{eq:supl_vargauge_discrete_t}
\end{align}
This expression ought to be compared with~\eqref{eq:supl_vargauge_discrete} for the gauge potential of time-dependent potential.

\section{Mean-field simulations of atomic transport}
\label{app:GPE}
\noindent We would also like to comment on the presence of atomic interactions in the data shown in the main text. While atomic interactions can play a non-trivial role in the dynamics of atoms in the momentum-space lattice (as demonstrated in Ref.~\cite{An-Inter}), for these experiments we operate in a regime in which the interactions do not significantly influence the results. Small quantitative effects, such as a slight asymmetry of the amount of transfer across the lattice in the ``lattice inversion experiment'' with respect to a positive or negative sweep of the lattice tilt, or small differences in the widths of the high and low-energy eigenstates in the ``eigenstate preparation experiment'' (which should be equivalent in the absence of interactions). These interaction effects were small, however, such that they did not lead to qualitative deviations from the single-particle behavior. Still, the calculated theory curves in Figs.~\pref[f]{CDtransfer}~and~\pref[g]{CDtransfer} take atomic interactions into account based on mean-field, Gross-Pitaevskii equation (GPE) calculations. These GPE calculations, which include interactions at roughly the same scale as the maximum tunneling strengths, provide an improved agreement with the observed ``lattice transfer'' data.

The interactions in our system of ultracold neutral atoms relate to short-ranged contact interactions in real space. In momentum space, these relate to long-ranged or all-to-all interactions between atoms occupying the discrete momentum orders. If these interactions were simply isotropic, or mode-independent, then the atomic interactions would be incapable of driving any correlated behavior, as the total interaction enegry would be a simple constant of motion. Indeed, the direct interactions between pairs of atoms occupying any set of momentum modes are independent of the mode occupation. However, there is an additional exchange energy for pairs of our identical bosonic atoms that interact while occupying distinguishable momentum states~\cite{Ozeri-RMP}. This added inter-mode interaction (repulsive in our case of having a positive scattering length for $^{87}$Rb), which is a consequence of bosonic statistics, \textit{i.e.}, the symmetry of the two-boson wave function, relates to an effectively attractive interaction between pairs of atoms occupying the same momentum mode~\cite{An-Inter}.

We model the particle dynamics within the synthetic lattice by the following mean-field equation:
\begin{equation}
    i \partial_\tau \psi(\tau) = H \psi(\tau) - U|\psi(\tau)|^2\psi(\tau) \ ,
\end{equation}
where $H$ is the single-particle tight-binding Hamiltonian (including any CD driving terms), $\psi(\tau)$ is the wave function with $\tau$ for time, and $U$ is the average mean-field interaction strength with the negative sign indicating an effectively attractive interaction. The atomic density of the BEC is not uniform but follows a Thomas--Fermi profile and therefore the mean-field strength should also be non-uniform. We account for this by taking a local-density approximation where the simulation is split into shells of roughly equivalent density and therefore equivalent mean-field interaction energy. These local simulations are then recombined by weighted averaging to yield the final result. In the case of Fig.~\pref[f]{CDtransfer} of the main text, the simulations were broken up into 20 shells based on an average mean-field interaction of $U/h = \unit[1.25]{kHz}$.

\end{document}